\documentclass[journal]{IEEEtran}

\usepackage{amsfonts}
\usepackage{amsmath}
\usepackage{amssymb}
\usepackage{balance}
\usepackage{color}
\usepackage{epsfig}
\usepackage{float}
\usepackage{graphics}
\usepackage{multirow}
\usepackage{rotating}
\usepackage{stfloats}
\usepackage{subfigure}
\usepackage{textcomp}
\usepackage{url}
\usepackage{vector}
\usepackage{wasysym}

\IEEEoverridecommandlockouts

\def\B{\mathcal B}
\def\C{\mathcal C}
\def\O{\mathcal O}

\def\bfB{\mathbf{B}}

\begin{document}

\title{Threshold Analysis of Non-Binary\\Spatially-Coupled LDPC Codes\\with Windowed Decoding}

\author{
    \IEEEauthorblockN{Lai Wei\IEEEauthorrefmark{1}\IEEEauthorrefmark{2},
    Toshiaki Koike-Akino\IEEEauthorrefmark{2},
    David G. M. Mitchell\IEEEauthorrefmark{1},
    Thomas E. Fuja\IEEEauthorrefmark{1},
    and Daniel J. Costello, Jr.\IEEEauthorrefmark{1}}

    \IEEEauthorblockA{
    \IEEEauthorrefmark{1}
    Department of Electrical Engineering, University of Notre Dame, Notre Dame, IN, U.S\\\{lwei1, david.mitchell, tfuja, dcostel1\}@nd.edu}

    \IEEEauthorblockA{
    \IEEEauthorrefmark{2}Mitsubishi Electric Research Laboratories (MERL), Cambridge, MA, U.S.\\\{wei, koike\}@merl.com}
}

\maketitle

%**************************************************************
%**************************************************************
%   Abstract
%**************************************************************
%**************************************************************
\begin{abstract}
\label{Abstract}
In this paper we study the iterative decoding threshold performance of non-binary spatially-coupled low-density parity-check (NB-SC-LDPC) code ensembles for both the binary erasure channel (BEC) and the binary-input additive white Gaussian noise channel (BIAWGNC), with particular emphasis on windowed decoding (WD). We consider both $(2,4)$-regular and $(3,6)$-regular NB-SC-LDPC code ensembles constructed using protographs and compute their thresholds using protograph versions of NB density evolution and NB extrinsic information transfer analysis. For these code ensembles, we show that WD of NB-SC-LDPC codes, which provides a significant decrease in latency and complexity compared to decoding across the entire parity-check matrix, results in a negligible decrease in the near-capacity performance for a sufficiently large window size $W$ on \emph{both} the BEC and the BIAWGNC. Also, we show that NB-SC-LDPC code ensembles exhibit gains in the WD threshold compared to the corresponding block code ensembles decoded across the entire parity-check matrix, and that the gains increase as the finite field size $q$ increases. Moreover, from the viewpoint of decoding complexity, we see that $(3,6)$-regular NB-SC-LDPC codes are particularly attractive due to the fact that they achieve near-capacity thresholds even for small $q$ and $W$.
\end{abstract}

%************************************************************************************************************
%************************************************************************************************************
\section{Introduction}
\label{Sec: Introduction}
Non-binary low-density parity-check (NB-LDPC) block codes constructed over finite fields of size $q>2$ outperform comparable binary LDPC block codes~\cite{LDPC_GFq}, in particular when the blocklength is short to moderate; however, this performance gain comes at the cost of an increase in decoding complexity. A direct implementation of the belief-propagation (BP) decoder~\cite{LDPC_GFq} has complexity $\O(q^2)$ per symbol. More recently, an implementation based on the fast Fourier transform~\cite{NB-LDPC-FFT} was shown to reduce the complexity to $\O(q \log q)$. Beyond that, a variety of simple but sub-optimal decoding algorithms have been proposed in the literature~\cite{EMS}~\cite{T-EMS}. As for computing iterative decoding thresholds, a non-binary extrinsic information transfer (NB-EXIT) analysis was proposed in~\cite{NB-EXIT} and was later developed into a corresponding version P-NB-EXIT~\cite{P-NB-EXIT} suitable for \emph{protograph}-based codes.

\begin{figure}[!t]
    \centerline{\includegraphics[width=0.6\columnwidth]{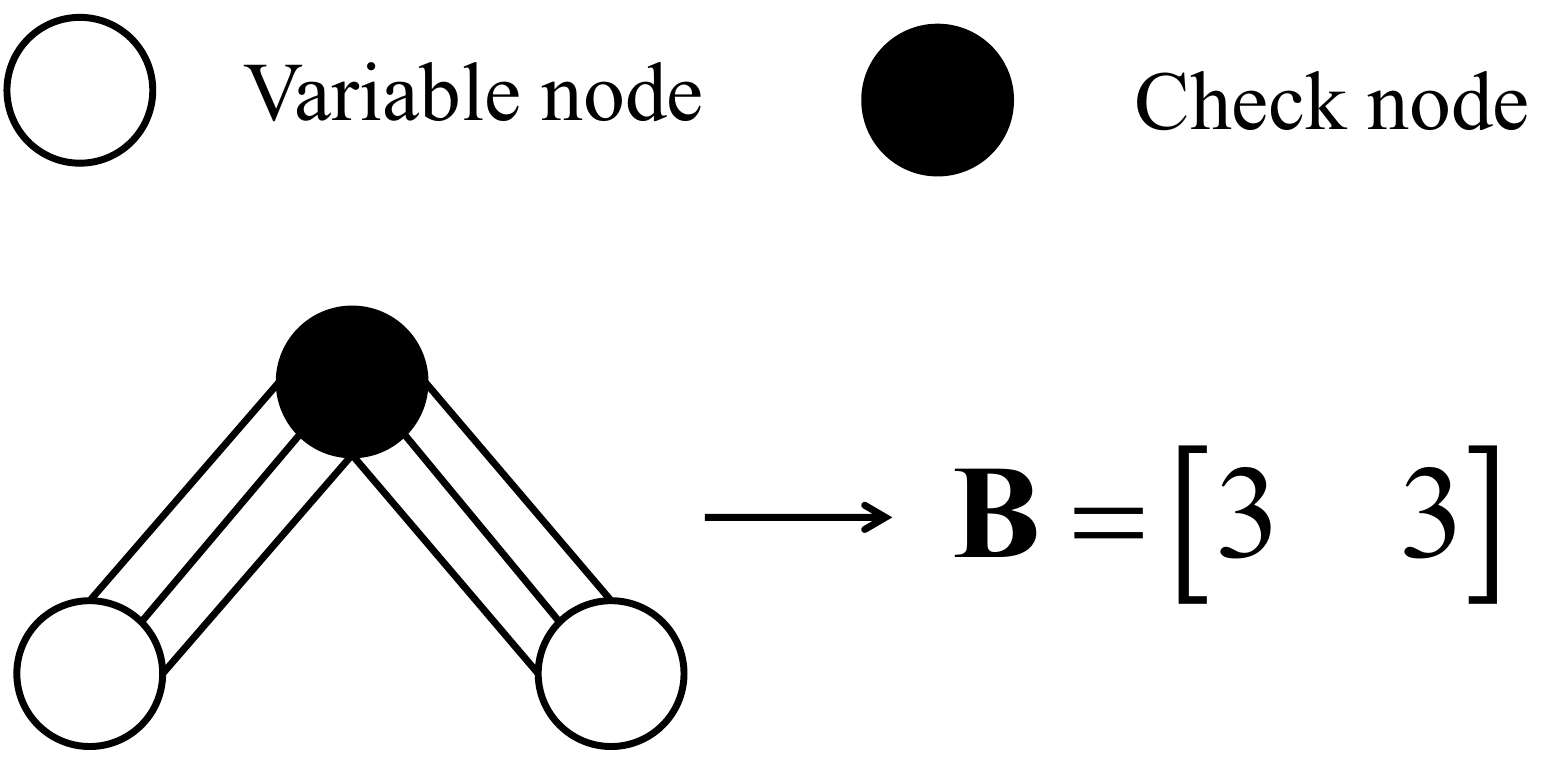}}
    \caption{A $(3,6)$-regular protograph and its corresponding base-matrix representation.}
    \label{Fig: protograph}
\end{figure}

A protograph~\cite{protograph} is a small Tanner graph, which can be used to produce a \emph{structured} LDPC code ensemble by applying a graph lifting procedure, such that every code in the ensemble maintains the structure of the protograph, i.e., it has the same degree distribution and the same type of edge connections. Figure~\ref{Fig: protograph} illustrates a $(3,6)$-regular protograph, which can be used to produce a $(3,6)$-regular LDPC block code ensemble. A protograph with $(c-b)$ check nodes and $c$ variable nodes can be represented equivalently by a \emph{base (parity-check) matrix} $\bfB$ consisting of non-negative integers, in which the $(i,j)$-th entry ($1 \leq i \leq c-b$ and $1 \leq j \leq c$) is the number of edges between check node $i$ and variable node $j$. To calculate the BP threshold of a protograph-based code ensemble, conventional tools must be adapted to take the edge connections into account. Although some freedom is lost in the code design when the protograph structure is adopted, one can use these modified protograph-based analysis tools to find ``good'' protographs with better BP thresholds than corresponding unstructured ensembles with the same degree distribution.
%************************************************************************
%************************************************************************
%************************************************************************
\newcounter{tempequationcounter}
\begin{figure*}[!b]
\normalsize
\setcounter{tempequationcounter}{\value{equation}}
\hrulefill
\begin{IEEEeqnarray}{rCl}
\setcounter{equation}{1}
    \bfB^\intercal_{\text{SC}} =
    \left[
        \begin{array}{c c c c c c c c c c}
        \bfB_0^\intercal & \bfB_1^\intercal & \dots & \bfB_{m_s}^\intercal & & & & & & \\
         & \bfB_0^\intercal & \bfB_1^\intercal & \dots & \bfB_{m_s}^\intercal & & & & & \\
         & & \dots & \dots & \dots & \dots & & & & \\
         & & & & & & \bfB_0^\intercal & \bfB_1^\intercal & \dots & \bfB_{m_s}^\intercal \\
        \end{array}
    \right]_{c L \times (c-b) (L+m_s)}
\label{Eq: SC protograph}
\end{IEEEeqnarray}
\setcounter{equation}{\value{tempequationcounter}}
\end{figure*}
%************************************************************************

Spatially-coupled LDPC (SC-LDPC) codes, also known as terminated LDPC convolutional codes~\cite{terminated-LDPC}, have been shown to exhibit a phenomenon called ``threshold saturation''~\cite{threshold-saturation}, where, as the termination length grows, the BP decoding threshold saturates to the maximum \emph{a-posteriori} (MAP) threshold of a $(d_v,d_c)$-regular underlying LDPC block code ensemble, which in turn improves to the channel capacity as the density ($d_v$ and $d_c$) of the parity-check matrix increases. Iterative decoding threshold results on the binary erasure channel (BEC) for non-binary SC-LDPC (NB-SC-LDPC) code ensembles have been reported by Uchikawa~\emph{et al.}~\cite{RC-NB-LDPCC} and Piemontese~\emph{et al.}~\cite{NBSC-BEC}, and the corresponding threshold saturation was proved by Andriyanova~\emph{et al.}~\cite{NBSC-th-sat}. In each of these papers, the authors assumed that decoding was carried out across the entire parity-check matrix of the code; for simplicity, this will be referred to as the flooding schedule (FS) in our paper.

A major problem with FS decoding of SC-LDPC codes is latency. To resolve this issue, a more efficient technique, called windowed decoding (WD), was proposed in~\cite{WD-Iyengar}. Compared to FS decoding, WD exploits the convolutional nature of the SC parity-check matrix to localize the decoder and thereby reduce latency; under WD, the decoding window contains only a portion of the parity-check matrix, and within that window BP decoding is performed.

In this paper, assuming that the binary image of a codeword is transmitted, we analyze the WD threshold performance of $(2,4)$-regular and $(3,6)$-regular NB-SC-LDPC code ensembles based on protographs. In particular,
\begin{enumerate}
    \item for the BEC, we develop the NB density evolution (NB-DE) analysis as proposed in~\cite{NB-DE-BEC} into a protograph version, which we call P-NB-DE, and
    \item for the binary-input additive white Gaussian noise channel (BIAWGNC) with binary phase-shift keying (BPSK) modulation, we apply the P-NB-EXIT~\cite{P-NB-EXIT} analysis (originally proposed for NB-LDPC block codes) to NB-SC-LDPC codes.
\end{enumerate}
The finite field size $q$ is constrained to be $2^m$, where $m$ is a positive integer.

In both cases, our primary contribution lies in the scenario when WD is implemented, so that decoder latency can be reduced at the cost of a small loss in decoder performance. For three NB-SC-LDPC ensemble examples, we show in Sections~\ref{Sec: Th BEC} and~\ref{Sec: Th AWGN} that WD provides two of the ensembles with non-decreasing threshold performance as $W$ and/or $m$ increases. In fact, their WD thresholds are numerically capacity-achieving for sufficiently large $W$ and $m$. As for the third ensemble, although its WD threshold diverges slightly from capacity when $m$ is large (observed on the BEC), it is the strongest candidate for low-latency and/or low-complexity applications due to its excellent performance when $W$ and $m$ are \emph{both} small; this conclusion is further strengthened in our analysis of WD complexity in Section~\ref{Sec: Dec cpl}. In all, the results of this paper provide theoretical guidance for designing and implementing practical NB-SC-LDPC codes for WD.

%************************************************************************************************************
%************************************************************************************************************
\section{NB-SC-LDPC Code Ensembles}
\label{Sec: NBSC}
An SC-LDPC code ensemble can be constructed from a LDPC block code ensemble using an edge spreading technique~\cite{edge-spread}, which can be described conveniently by protographs.

As shown in Figure~\ref{Fig: protograph}, let $\bfB$ denote a block base matrix of size $(c-b) \times c$, which corresponds to a protograph representation of an LDPC block code ensemble with design rate $R=b/c$. An SC base matrix corresponding to an SC-LDPC code ensemble can then be constructed using $(m_s+1)$ component base matrices $\{\bfB_i\}_{i=0}^{m_s}$, each of size $(c-b) \times c$, where the edges of $\bfB$ are spread such that
\begin{eqnarray*}
    \sum_{i=0}^{m_s} \bfB_i = \bfB,
\end{eqnarray*}
and $m_s$ is the memory size. The resulting SC base matrix is given in its transpose form in (\ref{Fig: protograph}) at the bottom of this page, where $L$ is called the termination length. The design rate of the code is
\begin{eqnarray*}
    R_L = 1 - \dfrac{(c-b) (L+m_s)}{c L}.
\end{eqnarray*}
As a result of the termination, there is a rate loss compared to the block code design rate; however, this diminishes as $L$ increases, i.e., $R_L \rightarrow R = b/c$ when $L \rightarrow \infty$. In WD, the window size $W$ is defined as the number of \emph{column blocks} of size $c$ covered by the decoding window, which slides over a portion of $\bfB_{\text{SC}}$ of fixed size $W(c-b)$ by $Wc$ (in symbols, see~\cite{WD-Iyengar} for details).

In this paper, we use the following three protographs as examples, where $\C$ denotes the SC-LDPC code ensembles and $\B$ denotes the underlying LDPC block code ensembles:
\begin{enumerate}
    \item $\B_{[2,4]}$ and $\C_{[2,4]}$: The block base matrix $\bfB$ representing a $(2,4)$-regular LDPC block code ensemble $\B_{[2,4]}$ and the component matrices used to construct an SC base matrix $\bfB_{\text{SC}}$ representing an SC-LDPC code ensemble $\C_{[2,4]}$, are given by
        \vspace{-0.5mm}
        \begin{eqnarray*}
            &\bfB =
            \left[
                \begin{array}{c c}
                2 & 2 \\
                \end{array}
            \right] \Rightarrow
            \bfB_0 = \bfB_1 =
            \left[
                \begin{array}{c c}
                1 & 1 \\
                \end{array}
            \right].
        \end{eqnarray*}
        As noted above, the value of an entry in $\bfB$ (resp. $\bfB_{\text{SC}}$) is equal to the number of edges connecting the corresponding check node and variable node in the protograph for $\B$ (resp. $\C$).
    \item $\B_{[3,6]}$ and $\C_{[3,6]}^{m_s=1}$: The block base matrix $\bfB$ corresponding to a $(3,6)$-regular LDPC block code ensemble $\B_{[3,6]}$ and the component matrices $\bfB_0$ and $\bfB_1$ corresponding to an SC-LDPC code ensemble $\C_{[3,6]}^{m_s=1}$ are given by
        \vspace{-0.5mm}
        \begin{eqnarray*}
            \bfB =
            \left[
                \begin{array}{c c}
                3 & 3 \\
                \end{array}
            \right] \Rightarrow
            \bfB_0 =
            \left[
                \begin{array}{c c}
                2 & 1 \\
                \end{array}
            \right],
            \bfB_1 =
            \left[
                \begin{array}{c c}
                1 & 2 \\
                \end{array}
            \right].
        \end{eqnarray*}
    \item $\C_{[3,6]}^{m_s=2}$:
        \vspace{-0.5mm}
        \begin{eqnarray*}
            \bfB =
            \left[
                \begin{array}{c c}
                3 & 3 \\
                \end{array}
            \right] \Rightarrow
            \bfB_0 = \bfB_1 = \bfB_2 =
            \left[
                \begin{array}{c c}
                1 & 1 \\
                \end{array}
            \right].
        \end{eqnarray*}
\end{enumerate}
For each example, the termination length is chosen as $L = 100$, so that $R_L$ is close to $R$. We will refer to the ``$(2,4)$ group'' as the collection of ensembles $\B_{[2,4]}$ and $\C_{[2,4]}$ and the ``$(3,6)$ group'' as $\B_{[3,6]}$, $\C_{[3,6]}^{m_s=1}$, and $\C_{[3,6]}^{m_s=2}$.

In practice, an NB-SC-LDPC code is generated from $\bfB_{\text{SC}}$ in two steps, similar to the procedure for generating an NB-LDPC block code from $\bfB$~\cite{protograph}:
\begin{enumerate}
    \item ``Lifting'': Replace the nonzero entries in $\bfB_{\text{SC}}$ by an $M \times M$ permutation matrix (or a sum of non-overlapping $M \times M$ permutation matrices), and replace the zero entries by the $M \times M$ all-zero matrix; $M$ is called the lifting factor. In this way, the structure of $\bfB_{\text{SC}}$ is maintained in the lifted SC-LDPC matrix, so the threshold analysis of the SC-LDPC code ensemble $\C$ can be carried out directly based on $\bfB_{\text{SC}}$.
    \item ``Labeling'': Randomly assign to each non-zero entry in the lifted parity-check matrix a non-zero element in $\text{GF}(q)$, where $q = 2^{m}$ is the finite field size.
\end{enumerate}
After the lifting step, the parity-check matrix is still binary, i.e., the non-binary feature does not arise until labeling. Both the permutation matrices and the selection of labels can be optimized in order to obtain a good code~\cite{P-NB-EXIT}, but this is not our emphasis here, since we are interested in a threshold analysis of the general non-binary ensemble, where the dimension of the message model used in the threshold analysis depends on the size of the finite field~\cite{NB-EXIT}~\cite{NB-DE-BEC}.

%************************************************************************************************************
%************************************************************************************************************
\section{Threshold Analysis of \\NB-SC-LDPC Code Ensembles on the BEC}
\label{Sec: Th BEC}
\subsection{P-NB-DE Analysis on the BEC}
\label{Sec: Th BEC; Subsec: P-NBDE}
We extended the NB-DE algorithm for the BEC~\cite{NB-DE-BEC} to a protograph version, which we denote P-NB-DE, similar to the procedure used to extend NB-EXIT to P-NB-EXIT in~\cite{P-NB-EXIT}. Since edge connections are taken into account, P-NB-DE is essentially the BP algorithm performed on a protograph. The resulting BP threshold is denoted $\epsilon^{\text{BP}}(\delta)$ with $\delta \in [0,1]$, which is the largest channel erasure rate such that all transmitted symbols can be recovered successfully with probability at least $(1-\delta)$ when the number of iterations goes to infinity. Our results are obtained for $\delta = 10^{-6}$; however, we note that setting $\delta=10^{-12}$ provides similar results.

\subsection{Numerical Results}
\label{Sec: Th BEC; Subsec: Sim}
\begin{figure}[!t]
    \centering
    \includegraphics[width=0.75\columnwidth]{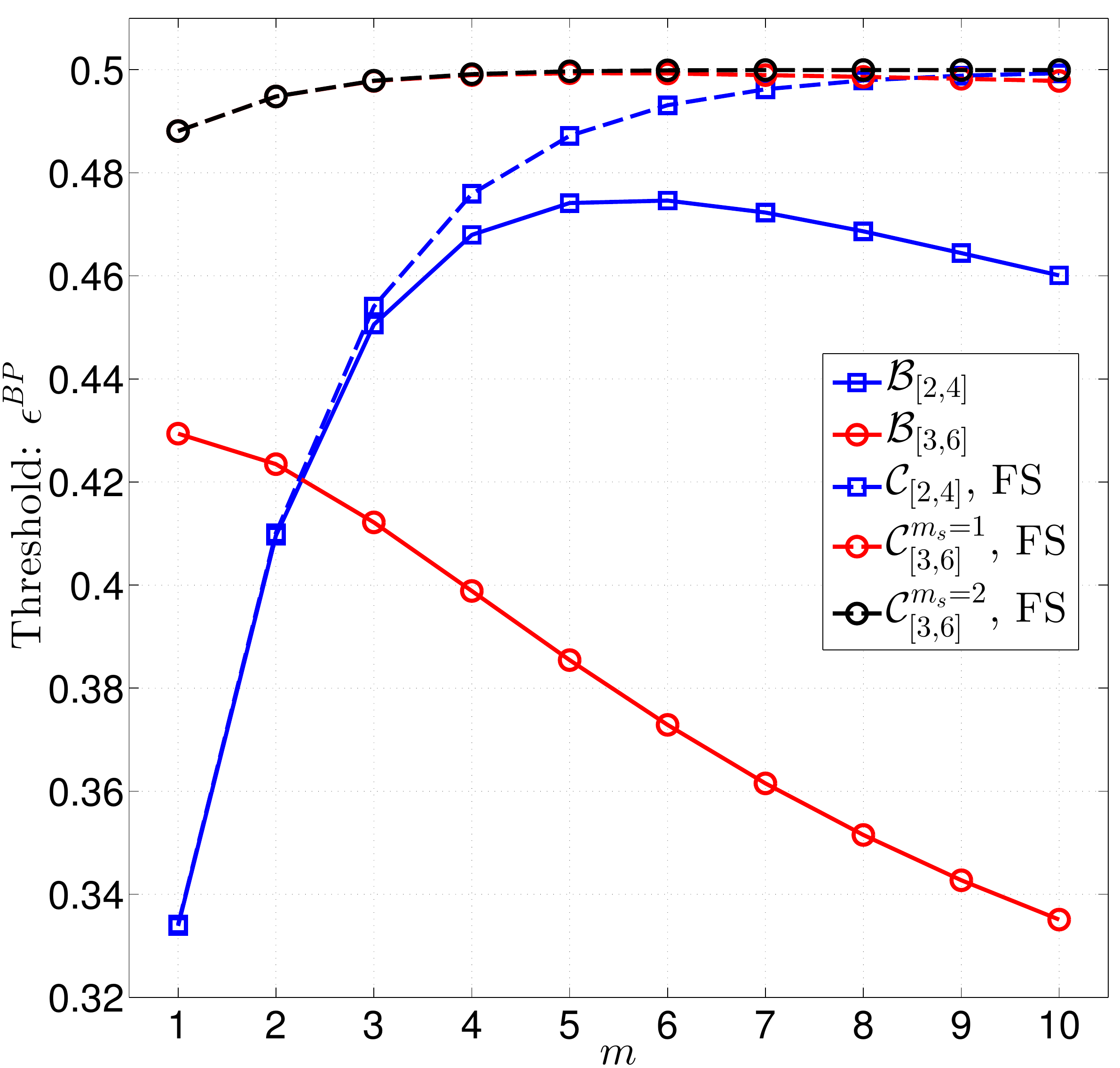}
    \caption{Comparison of the thresholds based on the flooding schedule (FS) for the $(2,4)$ and $(3,6)$ groups on the BEC.}
    \label{Fig: TH BEC FS}
\end{figure}

In this section, we present the numerical results for the BEC, with emphasis on the threshold performance of NB-SC-LDPC code ensembles when WD is used. As a benchmark, Figure~\ref{Fig: TH BEC FS} first compares the FS threshold performance of the $(2,4)$ group and the $(3,6)$ group, where FS decoding is carried out on the entire parity-check matrix and not restricted to a window. We observe that the NB-SC-LDPC codes perform extremely well compared to their block code counterparts, in particular for large field size $m$. Unlike the block code case, $\C_{[3,6]}^{m_s=2}$ always outperforms $\C_{[2,4]}$, and the thresholds of these two ensembles increase monotonically with $m$. However, this monotonic increase is not observed for $\C_{[3,6]}^{m_s=1}$, where the obtained threshold actually decreases very slightly for $m>5$. (See the related discussion of Figure~\ref{Fig: TH BEC 3 6 ms 1} below.)
%************************************************************************
%************************************************************************

Figure~\ref{Fig: TH BEC 2 4} shows that, for a sufficiently large window size $W$, WD provides threshold performance nearly the same as the FS, i.e., the performance loss is negligible while the decoder benefits from greatly reduced delay. We define $W^{*}$ to be the smallest window size such that WD provides a threshold within $3\%$ of the FS threshold universally for \emph{all} field sizes $m$.\footnote{This ``$3\%$'' value is actually loose for moderate to large $m$. In the cases we examined, the WD threshold with $W=W^{*}$ typically lies within $0.5\%$ of or is even numerically identical to the FS threshold for $m>2$.} For $\C_{[2,4]}$, we find $W^{*} = 30$.
%************************************************************************
\begin{figure}[!t]
    \centering
    \subfigure[$\B_{[2,4]}$ and $\C_{[2,4]}$]
    {
        \includegraphics[width=0.75\columnwidth]{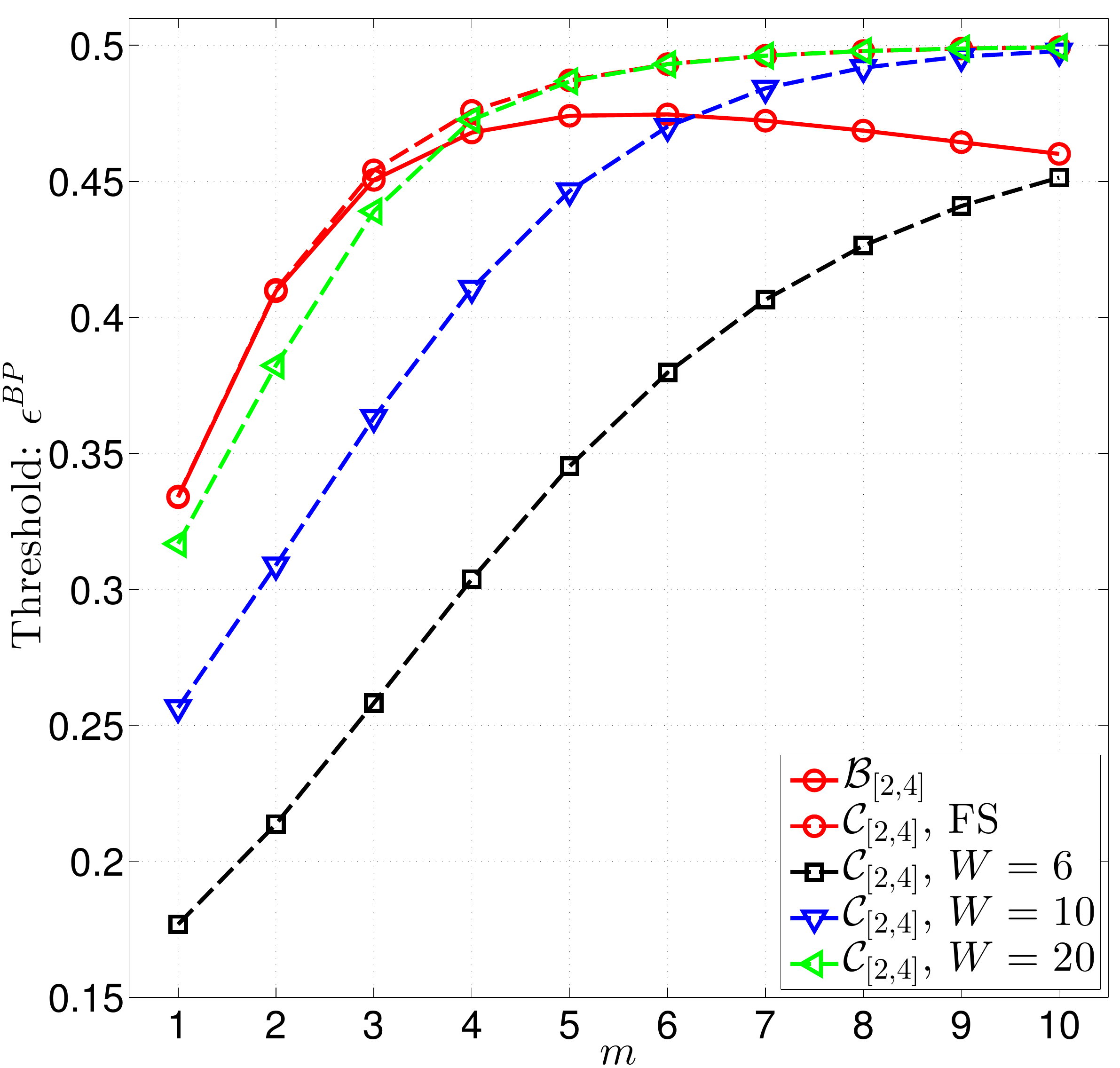}
        \label{Fig: TH BEC 2 4}
    }
    \subfigure[$\B_{[3,6]}$ and $\C_{[3,6]}^{m_s=1}$]
    {
        \includegraphics[width=0.75\columnwidth]{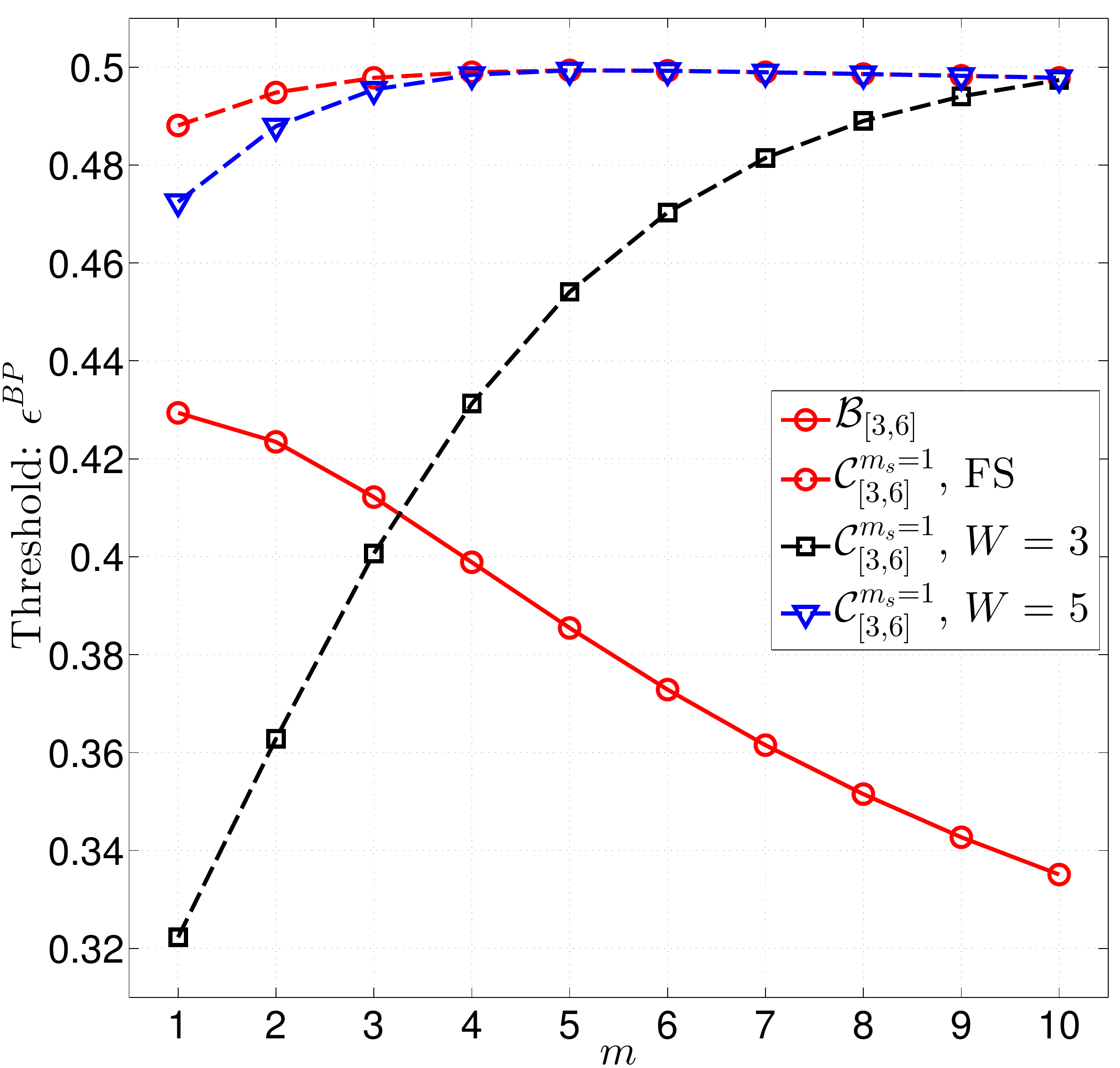}
        \label{Fig: TH BEC 3 6 ms 1}
    }
    \subfigure[$\B_{[3,6]}$ and $\C_{[3,6]}^{m_s=2}$]
    {
        \includegraphics[width=0.75\columnwidth]{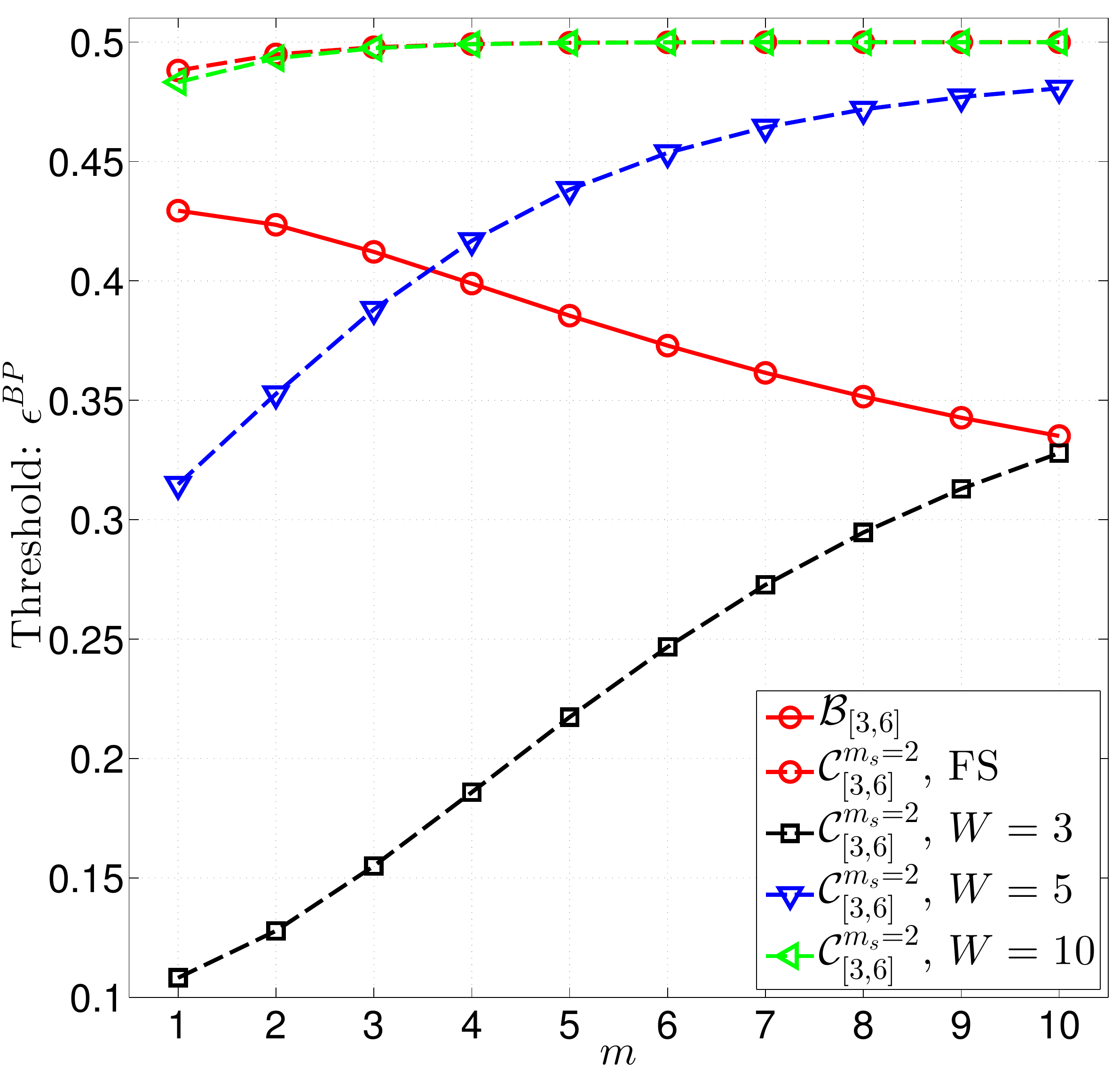}
        \label{Fig: TH BEC 3 6 ms 2}
    }
    \caption{FS and WD thresholds of the $(2,4)$ and $(3,6)$ groups on the BEC.}
    \label{Fig: TH BEC WD}
\end{figure}
%************************************************************************

Similar observations can be made for WD of $\C_{[3,6]}^{m_s=1}$ and $\C_{[3,6]}^{m_s=2}$ in Figure~\ref{Fig: TH BEC 3 6 ms 1} and~\ref{Fig: TH BEC 3 6 ms 2}:
\begin{itemize}
    \item For $\C_{[3,6]}^{m_s=1}$, $W^{*} = 7$. The WD threshold grows to a value within $0.1\%$ of the channel capacity when $m=5$, and then decreases very slightly as $m$ increases further. Nevertheless, the WD threshold remains very close to capacity even for large $m$.
    \item For $\C_{[3,6]}^{m_s=2}$, $W^{*} = 10$. The WD threshold does not degrade as $m$ increases, but instead saturates to a value numerically indistinguishable from capacity. However, when $W$ is small (e.g., $W=5$), $\C_{[3,6]}^{m_s=2}$ does not perform as well as $\C_{[3,6]}^{m_s=1}$ due to its larger memory size, which increases the delay required to make reliable decisions (see~\cite{WD-Iyengar}).
\end{itemize}

To summarize, for the three considered NB-SC-LDPC code ensembles, the gain introduced by spatial coupling compared to the corresponding uncoupled NB-LDPC block code ensembles grows with increasing field size $m$ for sufficiently large window size $W$. Furthermore, the thresholds of the $\C_{[2,4]}$ and $\C_{[3,6]}^{m_s=2}$ NB-SC-LDPC code ensembles saturate to a value numerically indistinguishable from capacity for large $m$ with either FS decoding or WD with a sufficiently large $W$. This is analogous to the case where the thresholds of binary $(d_v,d_c)$-regular SC-LDPC code ensembles saturate to capacity as $d_v$ and $d_c$ get large. It is interesting to note that for binary ensembles the graph density ($d_v$ and $d_c$) must get large to approach capacity, whereas for non-binary ensembles capacity can be approached for fixed density by increasing the field size. In other words, the increase in complexity needed to approach capacity is different in the two cases.

%************************************************************************************************************
%************************************************************************************************************
\section{Threshold Analysis of\\NB-SC-LDPC Code Ensembles on the BIAWGNC}
\label{Sec: Th AWGN}
\subsection{P-NB-EXIT Analysis on the BIAWGNC}
\label{Sec: Th AWGN; Subsec: P-EXIT}
%************************************************************************
\begin{figure}[!t]
    \centering
    \subfigure[Comparison of the $(2,4)$ and $(3,6)$ groups: FS]
    {
        \includegraphics[width=0.75\columnwidth]{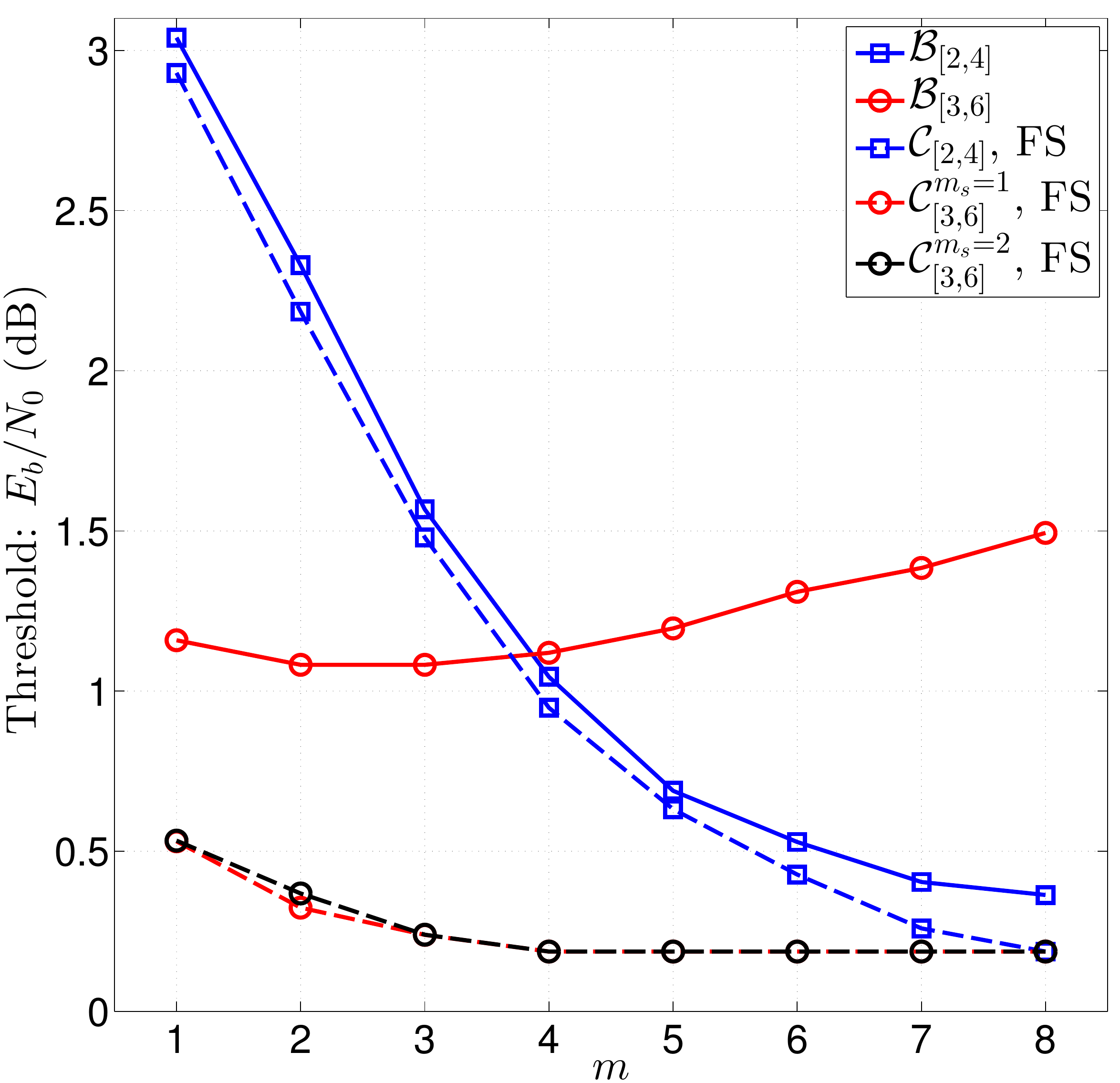}
        \label{Fig: TH AWGN FS}
    }
    \subfigure[$\B_{[3,6]}$ and $\C_{[3,6]}^{m_s=1}$]
    {
        \includegraphics[width=0.75\columnwidth]{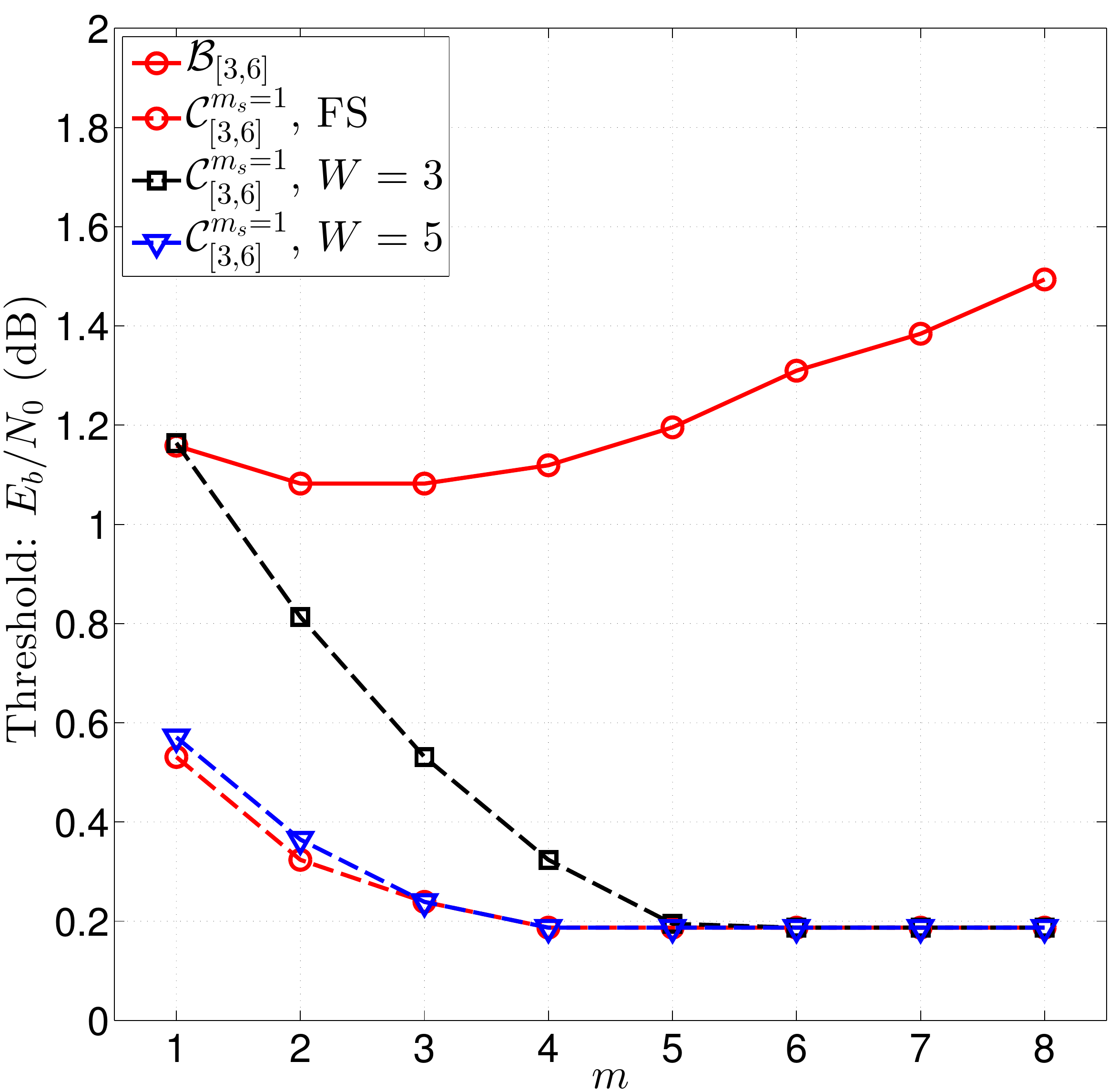}
        \label{Fig: TH AWGN 3 6 ms 1}
    }
    \caption{FS thresholds of the $(2,4)$ and $(3,6)$ groups and WD thresholds of $\C_{[3,6]}^{m_s=1}$ on the BIAWGNC.}
    \label{Fig: TH AWGN}
\end{figure}
%************************************************************************
We use the P-NB-EXIT algorithm presented in~\cite{P-NB-EXIT} to analyze the threshold performance of NB-SC-LDPC code ensembles on the BIAWGNC, assuming that the binary image of a codeword is transmitted and that BPSK modulation is used. Similar to the P-NB-DE analysis on the BEC, the P-NB-EXIT analysis is also a BP algorithm performed on the protograph, where the messages represent \emph{mutual information} (MI) values, a model obtained by approximating the distribution of the log-likelihood ratios as (jointly) Gaussian. The threshold is obtained by determining the smallest signal-to-noise ratio $E_b/N_0$ such that decoding is successful, i.e., the smallest value of $E_b/N_0$ such that the \emph{a-posteriori} MI between each variable node and a corresponding codeword symbol goes to $1$ as the number of iterations increases.

\subsection{Numerical Results}
\label{Sec: Th AWGN; Subsec: Sim}
Figure~\ref{Fig: TH AWGN FS} compares the FS thresholds of the $(2,4)$ and $(3,6)$ groups on the BIAWGNC and Figure~\ref{Fig: TH AWGN 3 6 ms 1} shows the WD thresholds of $\C_{[3,6]}^{m_s=1}$ for different $W$.\footnote{Due to computational complexity, the BIAWGNC thresholds were calculated only up to $m=8$. However, similar to the approach taken by Uchikawa \emph{et al.} in~\cite{RC-NB-LDPCC}, the BIAWGNC threshold performance for $m=9$ and $10$ is conjectured to be consistent with the corresponding BEC results.} Both figures illustrate similar behavior as the BEC results presented in Section~\ref{Sec: Th BEC; Subsec: Sim}, and the same is true for the WD thresholds of $\C_{[2,4]}$ and $\C_{[3,6]}^{m_s=2}$ (not included in the figure due to space limitations). To summarize, small gains are observed for $\C_{[2,4]}$ compared to $\B_{[2,4]}$ until the field size $m$ gets large, whereas numerically capacity-achieving WD thresholds that are significantly better than the corresponding block code thresholds are observed for both $\C_{[3,6]}^{m_s=1}$ and $\C_{[3,6]}^{m_s=2}$. Moreover, we find that $W^{*}=10$ for $\C_{[2,4]}$ and $\C_{[3,6]}^{m_s=2}$, while $\C_{[3,6]}^{m_s=1}$ is a better choice for WD, since $W^{*}=7$.

%************************************************************************************************************
%************************************************************************************************************
\section{Decoding Complexity}
\label{Sec: Dec cpl}
In practice, we would like to compare the performance of NB-SC-LDPC codes and the corresponding NB-LDPC block codes when their decoding latency is the same. Since it is assumed that the binary image of a codeword is transmitted, it is convenient to measure the latency in terms of $bits$, denoted as $W_b$, which is the number of columns in the window for WD of SC-LDPC codes and the blocklength of LDPC block codes, both measured in bits (instead of $GF(q)$ symbols). To be more specific, if the lifting factor (see Section~\ref{Sec: NBSC}) is $M$ for SC-LDPC codes and $M'$ for LDPC block codes, then the equal latency condition is given by
\begin{eqnarray*}
    W_b = Wc \cdot M \cdot m = c \cdot M' \cdot m,
\end{eqnarray*}
i.e., $M' = WM$, which means that SC-LDPC codes must use permutation matrices $W$ times smaller than LDPC block codes to maintain the same latency, where $c$ is the number of columns in the $\bfB$ and $\bfB_i$ matrices. For the codes we considered from the $(2,4)$ and $(3,6)$ groups, $c=2$. For fixed $M$, $W_b$ then depends on $Wm$. (The threshold analysis corresponds to the case when $M \to \infty$.)

As stated in~\cite{EMS} and the references therein, for NB-LDPC codes, if the BP algorithm employs the fast Fourier transform, then the computational complexity at a check node is $\O(qm) = \O(q\log_2q)$ \emph{per symbol} while that at a variable node is $\O(q)$. In our case, however, due to the constraint of equal latency, the decoding complexity should be estimated \emph{per window} for an SC-LDPC code, or equivalently, \emph{per blocklength} for an LDPC block code.

Like the protograph examples in this paper, an SC-LDPC code is typically derived from a $(d_v, d_c)$-regular LDPC block code. Consequently, if the window size is moderate to large, the part of the SC parity-check matrix covered by the window can be considered as $(d_v, d_c)$-regular as well and thus has (approximately) the same number of non-zero entries as the parity-check matrix of an LDPC block code. This indicates that the decoding complexity of an SC-LDPC code and an LDPC block code is the same when the decoding latency is the same.\footnote{In fact, due to the check-node irregularity at the beginning of the window and the variable-node irregularity at the end of the window, the actual decoding complexity of the SC-LDPC code is slightly lower than the LDPC block code. Nevertheless, we keep this ``regularity'' assumption for simplicity. Other factors that influence the decoding complexity, such as the number of iterations, are not considered.} The total number of non-zero entries in the window is $d_v W_b / m$, so the decoding complexity per window is
\begin{IEEEeqnarray}{rCl}
%\vspace{-1mm}
\setcounter{equation}{2}
    \O \left(
    \dfrac{W_b}{m}d_v\left( q + qm \right)
    \right) =
    M \O \left(
    W \cdot c \cdot d_v \left( q + qm \right)
    \right).
\label{Eq: WD dec cpl}
\end{IEEEeqnarray}

\begin{figure}[!t]
    \centerline{\includegraphics[width=\columnwidth]{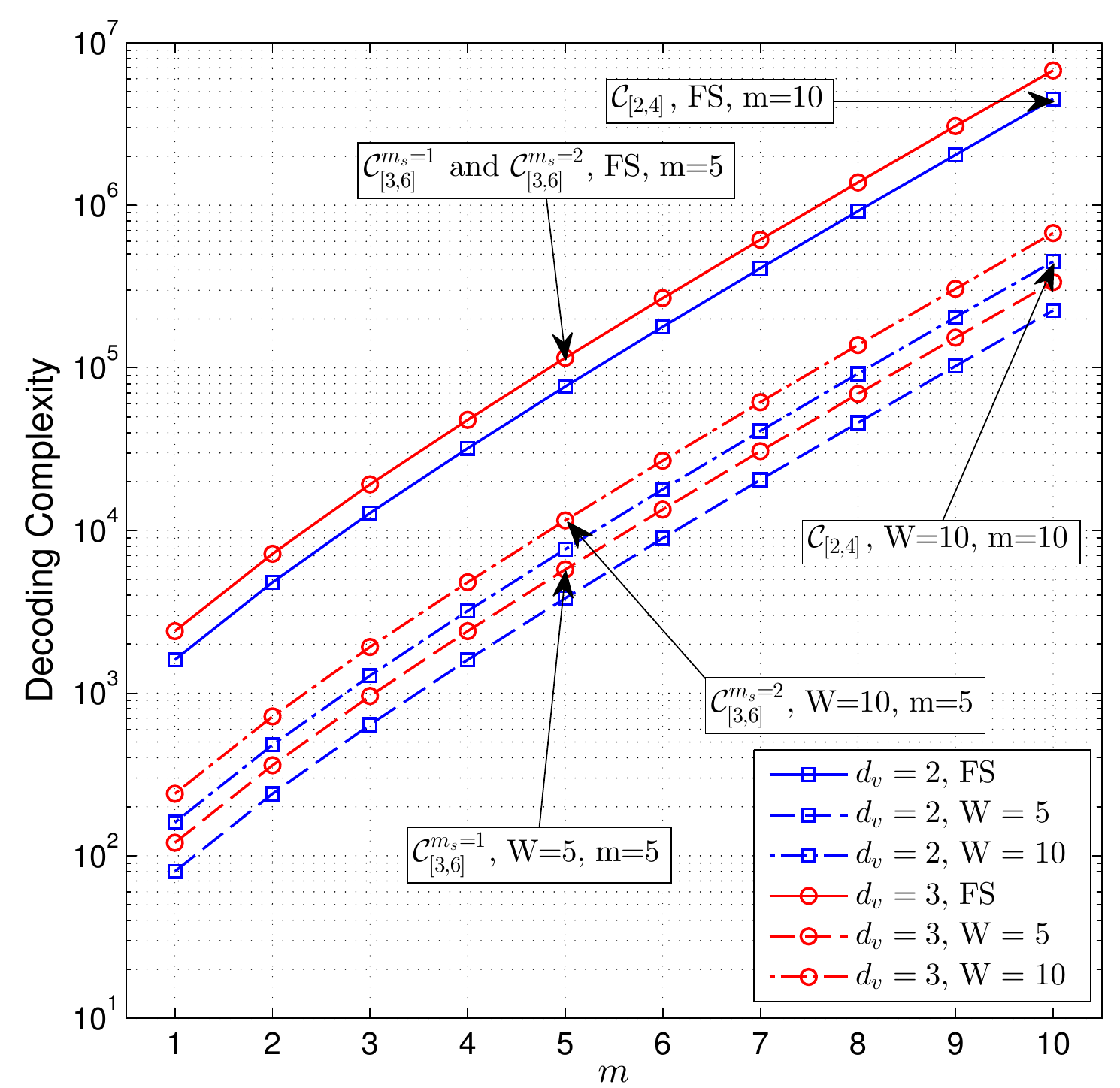}}
    \caption{The order of decoding complexity $\O \left(W \cdot c \cdot d_v \left( q + qm \right) \right)$ when an LDPC block code and an SC-LDPC code have the same decoding latency and thus have the same decoding complexity.}
    \label{Fig: dec cpl}
\end{figure}

Ignoring the $M$ factor on the right-hand side of (\ref{Eq: WD dec cpl}), Figure~\ref{Fig: dec cpl} shows the order of the decoding complexity when $d_v=2$ and $3$ with FS decoding and WD ($W=5$ and $10$); note that FS decoding is equivalent to WD with $W=L+m_s \approx L=100$, i.e., FS decoding corresponds to an increase in the window size by approximately an order of magnitude and to a corresponding order-of-magnitude increase in decoding complexity. The five specific points highlighted in the figure are all cases when FS decoding or WD threshold of an SC-LDPC code ensemble numerically achieves (or is very close to) capacity (recall that the FS threshold of an LDPC block code ensemble \emph{cannot} be capacity-achieving, as shown in Figures~\ref{Fig: TH BEC FS} and~\ref{Fig: TH AWGN FS}). We observe that
\begin{itemize}
    \item Comparing FS decoding to WD for the same ensemble, both the complexity and the latency are significantly reduced by adopting the latter.
    \item Comparing $\C_{[2,4]}$ ($Wm=100$), $\C_{[3,6]}^{m_s=2}$ ($Wm=50$), and $\C_{[3,6]}^{m_s=1}$ ($Wm=25$) for WD, \emph{both} the complexity and the latency of $\C_{[3,6]}^{m_s=1}$ are less than the other two ensembles for the same performance (near-capacity).
\end{itemize}
%************************************************************************
%************************************************************************

As a result, the $(3,6)$-regular construction (especially $\C_{[3,6]}^{m_s=1}$) is better than the $(2,4)$-regular construction when designing an NB-SC-LDPC code with decoding latency and complexity constraints and stringent performance requirements. This result is supported by decoding performance simulations of finite-length codes (see~\cite{NBSC-fixed-latency}).

%************************************************************************************************************
%************************************************************************************************************
\section{Conclusions}
\label{Sec: Conclusions}
This paper analyzed the windowed decoding threshold performance of several ensembles of non-binary spatially coupled LDPC codes; this was done for both the binary erasure channel and the BPSK-modulated additive white Gaussian noise channel. It was observed that windowed decoding (with a sufficiently large window) provides the spatially-coupled codes with capacity-approaching performance as the field size grows.  Moreover, the gain compared to the corresponding block code ensembles increases as well. One particular ensemble of $(3,6)$-regular NB-SC-LDPC codes with memory size $m_s=1$ was shown to exhibit near-capacity performance even for relatively small field and window sizes, i.e., low decoding complexity and small decoding latency.

%************************************************************************************************************
%************************************************************************************************************


\begin{thebibliography}{1}
\label{Bibi}
\balance
\bibitem{LDPC_GFq} M.~C.~Davey and D.~J.~C.~MacKay, ``Low-density parity check codes over $GF(q)$,'' \emph{IEEE Commun. Letters,} vol.~2, no.~6, pp.~165-167, Jun. 1998.
\bibitem{NB-LDPC-FFT} L.~Barnault and D.~Declercq, ``Fast decoding algorithm for LDPC over GF($2^q$)," in \emph{Proc. IEEE Inf. Theory Workshop}, pp.~70-73, Paris, France, Apr. 2003.
\bibitem{EMS} A.~Voicila, D.~Declercq, F.~Verdier, M.~Fossorier, M., and P.~Urard, ``Low-complexity decoding for non-binary LDPC codes in high order fields,'' \emph{IEEE Trans. Commun.,} vol.~58, no.~5, pp.~1365-1375, May 2010.
\bibitem{T-EMS} Erbao Li, D.~Declercq, and K.~Gunnam, ``Trellis-based extended min-sum algorithm for non-binary LDPC codes and its hardware structure,'' \emph{IEEE Trans. Commun.,} vol.~61, no.~7, pp.~2600-2611, Jul. 2013.
\bibitem{NB-EXIT} A.~Bennatan and D.~Burshtein, ``Design and analysis of nonbinary LDPC codes for arbitrary discrete-memoryless channels,'' \emph{IEEE Trans. Inf. Theory,} vol.~52, no.~2, pp.~549-583, Feb. 2006.
\bibitem{P-NB-EXIT} L.~Dolecek, D.~Divsalar, Y.~Sun, and B.~Amiri, ``Non-binary protograph-based LDPC codes: Enumerators, analysis, and designs,'' 2013. [Online]. Available: http://www.seas.ucla.edu/csl/files/publications/
\bibitem{protograph} J. Thorpe, ``Low-density parity-check (LDPC) codes constructed from protographs,'' JPL IPN Progress Report 42-154, Aug. 2003.
\bibitem{terminated-LDPC} M.~Lentmaier, A.~Sridharan, D.~J.~Costello, Jr., and K.~Sh.~Zigangirov, ``Iterative decoding threshold analysis for LDPC convolutional codes,'' \emph{IEEE Trans. Inf. Theory}, vol~56, no.~10, pp.~5274-5289, Oct. 2010.
\bibitem{threshold-saturation} S.~Kudekar, T.~J.~Richardson, and R.~L.~Urbanke, ``Threshold saturation via spatial coupling: Why convolutional LDPC ensembles perform so well over the BEC,'' \emph{IEEE Trans. Inf. Theory,} vol.~57, no.~2, pp.~803-834, Feb. 2011.
\bibitem{RC-NB-LDPCC} H.~Uchikawa, K.~Kasai, and K.~Sakaniwa, ``Design and performance of rate-compatible non-binary LDPC convolutional codes,'' 2011. [Online]. Available: http://arxiv.org/pdf/1010.0060v2.pdf
\bibitem{NBSC-BEC} A.~Piemontese, A.~G.~Amat, and G.~Colavolpe, ``Nonbinary spatially-coupled LDPC codes on the binary erasure channel,'' in \emph{Proc. IEEE Int. Conf. Commun.,} pp.~3270-3274, Budapest, Hungary, Jun. 2013.
\bibitem{NBSC-th-sat} I.~Andriyanova and A.~G.~Amat, ``Threshold saturation for nonbinary SC-LDPC codes on the binary erasure channel,'' 2013. [Online]. Available: http://arxiv.org/abs/1311.2003/
\bibitem{WD-Iyengar} A.~R.~Iyengar, M.~Papaleo, P.~H.~Siegel, J.~K.~Wolf, A.~Vanelli-Coralli, and G.~E.~Corazza, ``Windowed decoding of protograph-based LDPC convolutional codes over erasure channels,'' \emph{IEEE Trans. Inf. Theory,} vol.~58, no.~4, pp.~2303-2320, Apr. 2012.
\bibitem{NB-DE-BEC} V.~Rathi and R.~L.~Urbanke, ``Density evolution, thresholds and the stability condition for non-binary LDPC codes,'' \emph{IEE Commun. Proc.,} vol.~152, no.~6, pp.~1069-1074, Dec. 2005.
\bibitem{edge-spread} M.~Lentmaier, G.~P.~Fettweis, K.~Sh.~Zigangirov, and D.~J.~Costello, Jr., ``Approaching capacity with asymptotically regular LDPC codes,'' in \emph{Proc. Inf. Theory and App. Workshop,} San Diego, CA, Feb. 2009.
\bibitem{Lentmaier2011} M.~Lentmaier, M.~M.~Prenda, and G.~P.~Fettweis, ``Efficient message passing scheduling for terminated LDPC convolutional codes,'' in \emph{Proc. IEEE Int. Symp. on Inf. Theory,} pp.~1826-1830, Saint Petersburg, Russia, Aug. 2011.
\bibitem{NBSC-fixed-latency} K.~Huang, D.~G.~M.~Mitchell, L.~Wei, X.~Ma, and D.~J.~Costello, Jr., ``Performance comparison of non-binary LDPC block and spatially coupled codes,'' submitted to \emph{IEEE Int. Symp. on Inf. Theory,} Jan. 2014.
\end{thebibliography}
\end{document}